# Time-correlated Window Carrier-phase Aided GNSS Positioning Using Factor Graph Optimization for Urban Positioning

Xiwei Bai, Weisong Wen*, and Li-Ta Hsu

*Abstract*— **This paper proposes an improved global navigation satellite system (GNSS) positioning method that explores the time correlation between consecutive epochs of the code and carrier phase measurements which significantly increases the robustness against outlier measurements. Instead of relying on the time difference carrier phase (TDCP) which only considers two neighboring epochs using an extended Kalman filter (EKF) estimator, this paper proposed to employ the carrier-phase measurements inside a window, the so-called window carrier-phase (WCP), to constrain the states inside a factor graph. A left null space matrix is employed to eliminate the shared unknown ambiguity variables and therefore, correlated the associated states inside the WCP. Then the pseudorange, Doppler, and the constructed WCP measurements are integrated simultaneously using factor graph optimization (FGO) to estimate the state of the GNSS receiver. We evaluated the performance of the proposed method in two typical urban canyons in Hong Kong, achieving the mean positioning error of 1.76 meters and 2.96 meters, respectively, using the automobile-level GNSS receiver. Meanwhile, the effectiveness of the proposed method is further evaluated using a low-cost smartphone level GNSS receiver and similar improvement is also obtained, compared with several existing GNSS positioning methods.**

## I. INTRODUCTION

The global navigation satellite system (GNSS) [1] is currently one of the major means of providing globally-referenced positioning for autonomous systems [2-4] with navigation requirements. With the increased availability of multiple satellite constellations, GNSS can provide satisfactory performance in open-sky areas [5] based on the received pseudorange measurements via single-point positioning (SPP). However, the positioning accuracy is significantly degraded in highly-urbanized cities such as Hong Kong, due to signal reflection caused by static buildings and dynamic objects [6] such as double-decker buses leading to the notorious non-light-of-sight (NLOS) receptions and multipath effects, so-called outlier measurements. To mitigate the effects of the GNSS outlier measurements, numerous methods were proposed by introducing additional information or sensors, such as 3D mapping aided GNSS [7-9], camera aided GNSS NLOS detection [10-12], and 3D LiDAR aided GNSS NLOS detection [6] or correction [10, 13]. Unfortunately, these methods rely on the availability of 3D mapping information or LiDAR sensors which is not suitable for low-cost applications.

The other research stream is to make use of the Doppler frequency or carrier-phase measurements to improve the robustness of the estimator against outlier measurements. The fusion of pseudorange and Doppler measurements [14] was studied using an extended Kalman filter (EKF). The Doppler measurement, which is less noisy compared with the pseudorange measurements, was employed to estimate the velocity of the GNSS receiver to further constrain the kinematic of the GNSS receiver together with the pseudorange measurements. Given the fact that the GNSS measurements are highly environmentally dependent and time-correlated [15], the conventional filtering-based method (e.g. EKF estimator) for GNSS positioning cannot simultaneously explore the time-correlation among historical measurements. As a result, the filtering-based estimator is still sensitive to unexpected outlier measurements. To fill this gap, our recent work in [16] proposed a factor graph optimization (FGO) based GNSS positioning method that integrates the pseudorange and Doppler frequency measurements, which make of the historical information to increase the robustness of the estimator. Improved accuracy is obtained in the evaluated datasets compared with the conventional EKF estimator. However, the potential of carrier-phase measurement with high accuracy was not explored in [16]. Compared with the pseudorange and the Doppler measurements, the carrier-phase is less sensitive to the multipath effects [1]. However, the received carrier-phase measurement involves the unknown integer ambiguity which should be resolved for high accuracy positioning which is commonly done in the GNSS real-time kinematic (GNSS-RTK) positioning. However, the GNSS-RTK requires additional continuous corrections from the nearby reference stations which is not usually available.

One alternative to using the carrier-phase is to make use of the time difference carrier phase (TDCP) [17-19] technique to eliminate the unknown integer ambiguity between the consecutive epochs. By differencing the carrier-phase measurements between two epochs based on the assumption that the integer ambiguity variables of the two consecutive epochs are the same, the derived TDCP displacements can be applied to constrain the position difference of the associated

Xiwei Bai, Weisong Wen and Li-Ta Hsu, are with Hong Kong Polytechnic University, Hong Kong (correspondence e-mail: welson.wen@polyu.edu.hk).

two epochs. Meanwhile, the TDCP displacements can also be used to estimate velocity by its derivation over time [17]. Considering that both the TDCP and the Doppler frequency can be applied to estimate the velocity, the team from the China University of Mining and Technology conducted continuous work to integrate both information to estimate the velocity of the GNSS receiver in [20-22]. Interestingly, the recent work by the team from the Chiba Institute of Technology proposed a reference station-free double-difference technique to remove the integer ambiguity variables across all the epochs [23]. However, all these methods only consider the TDCP measurements in two epochs. According to the nature of GNSS tracking loops, the integer ambiguity tends to be constant once the satellite signal is effectively tracked. In other words, the signals transmitted from the same satellite that tracked in multiple epochs share the same ambiguity. Interestingly, this is similar to the visual-inertial integrated navigation system [24] where the visual landmarks shared the same position (corresponding to the satellite ambiguity) in multiple epochs. In other words, the states in multiple epochs are strongly correlated by the visual landmarks tracked repetitively. To avoid the repetitive estimation of the position of the landmarks, the work in [24] proposed to eliminate the shared variable (position) of the landmarks from the visual observation function to, therefore, increase the computational efficiency and guarantee the exploration of the time-correlation simultaneously. Based on the similar elimination operation, the work in [25] proposed to avoid the integer ambiguity resolution in the integration of GNSS-RTK and inertial measurement unit (IMU). However, its performance relies on the cost of the applied IMU sensor. Inspired by the elimination operation work in [24] and our previous work in [16], this paper proposes to employ the carrier-phase measurements received from the same satellite inside a window, the so-called window carrier-phase (WCP), to constrain the states inside a factor graph. Meanwhile, since these carrier-phase measurements share the same integer ambiguity variables, a left null space matrix is employed to eliminate the ambiguity variables and therefore, correlate the associated states inside the window. To cope with the potential inconsistency in integer ambiguity variables, the M-estimator [26] is employed to mitigate the impacts of the cycle slips phenomenon. Then the factor graph is employed to fuse the proposed WCP constraint, pseudorange, and the Doppler measurements for GNSS positioning. The main contributions of this paper are listed as follows:

(1) This paper is a continuous work of [16] by extending the deployment of the high accuracy carrier-phase measurements via a window carrier-phase constraint, which enables exploration of the kinematic time-correlation between multiple epochs. Importantly, the conventional EKF estimator is not able to exploit the advantage of the proposed WCP since only the two consecutive epochs are considered due to its recursive form. Fortunately, the proposed factor graph-based formulation opens a window for the utilization of the high-accuracy carrier-phase measurements.

(2) This paper verifies the effectiveness of the proposed method with several challenging datasets collected in urban canyons of Hong Kong, via both automobile level and smartphone level GNSS receivers. Since the performance of the WCP is highly related to the size of the window, we analyze the impact of the window size on the performance of the proposed GNSS positioning. Moreover, the accommodation of the potential cycle slip, which can violate the assumption of shared integer ambiguity inside the WCP, is analyzed using the different setup of M-estimator.

The remainder of this paper is organized as follows. An overview of the proposed method is given in Section II. Section III presents the modeling of the pseudorange, Doppler, and conventional TDCP and their integration using the FGO. The derivation of the proposed WCP constraint is elaborated together with the application of WCP in FGO in Section IV. Several real experiments were performed to evaluate the effectiveness of the proposed method in Section V. Finally, conclusions are drawn, and future work is presented in Section VI.

## II. OVERVIEW OF THE PROPOSED METHOD

An overview of the proposed method in this paper is shown in Fig. 1. The inputs of the method are the raw pseudorange, Doppler frequency, and carrier-phase measurements from a GNSS receiver. The output is the state estimation of the GNSS receiver in the earth-centered, earth-fixed (ECEF) frame. The measurement modeling blocks remove the atmosphere and satellite clock errors involved in the raw measurements. Finally, three kinds of factors are integrated using FGO. In this paper, matrices are denoted as uppercase with bold letters. Vectors are denoted as lowercase with bold letters. Variable scalars are denoted as lowercase italic letters. Constant scalars are denoted as lowercase letters. Moreover, the GNSS positioning mentioned in this paper means the pseudorange and Doppler frequency measurement fusion using FGO based on our previous work in [16]. This paper focuses on employing the carrier phase to aid the GNSS positioning method.

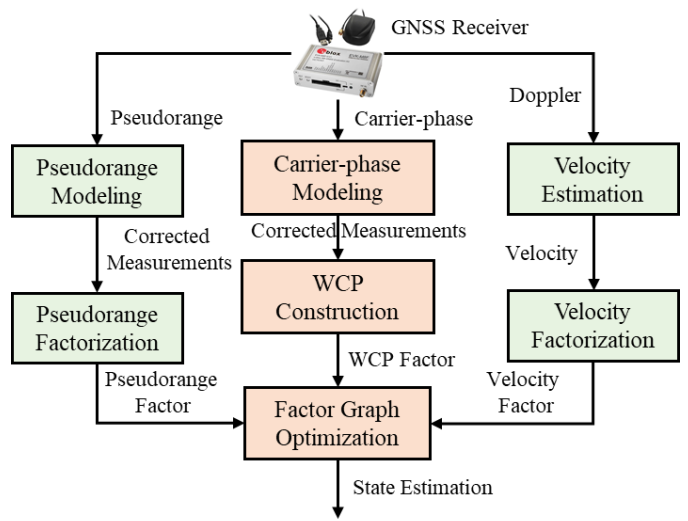

Fig. 1. Overview of the proposed method. The inputs are the continuous raw GNSS measurements. The output is the state estimation from the FGO.

To make the proposed pipeline clear, the following major notations are defined and followed by the rest of the paper.

a) The pseudorange measurement received from a satellite $s$ at a given epoch $t$ is expressed as $\rho_{r,t}^{s}$. The subscript $r$

denotes the GNSS receiver. The superscript $s$ denotes the index of the satellite.

b) The Doppler measurement received from satellite $s$ at a given epoch $t$ is expressed as $d_{r,t}^s$.

c) The carrier-phase measurement received from a satellite $s$ at a given epoch $t$ is expressed as $\psi_{r,t}^s$.

d) The position of the satellite $s$ at a given epoch $t$ is expressed as $\mathbf{p}_t^s = (p_{t,x}^s, p_{t,y}^s, p_{t,z}^s)^T$.

e) The velocity of the satellite $s$ at a given epoch $t$ is expressed as $\mathbf{v}_t^s = (v_{t,x}^s, v_{t,y}^s, v_{t,z}^s)^T$.

f) The position of the GNSS receiver at a given epoch $t$ is expressed as $\mathbf{p}_{r,t} = (p_{r,t,x}, p_{r,t,y}, p_{r,t,z})^T$.

g) The velocity of the GNSS receiver at a given epoch $t$ is expressed as $\mathbf{v}_{r,t} = (v_{r,t,x}, v_{r,t,y}, v_{r,t,z})^T$.

h) The clock bias of the GNSS receiver at a given epoch $t$ is expressed as $\delta_{r,t}$, that with the unit in meters. $\delta_{r,t}^s$ denotes the satellite clock bias by meters.

## III. GNSS Positioning Aided by TDCP Measurements

This section presents the methodology of the GNSS positioning aided by the TDCP measurement for further theoretical comparison between the conventional TDCP and the proposed WCP constraint (to be presented in Section IV). The utilization of TDCP is not original in the existing work using EKF [17], however, this paper still presents it for further theoretical and experimental comparison with the proposed WCP constraint. The state of the GNSS receiver is represented as follows:

$$\chi = [\mathbf{x}_{r,1}, \mathbf{x}_{r,2}, ..., \mathbf{x}_{r,n}] \qquad (1)$$

where the variable $\chi$ denotes the set of states of the GNSS receiver ranging from the first epoch to the current epoch $n$. The state of the GNSS receiver at a single epoch can be denoted as follows:

$$\mathbf{x}_{r,t} = (\mathbf{p}_{r,t}, \mathbf{v}_{r,t}, \delta_{r,t})^T \qquad (2)$$

where the $\mathbf{x}_{r,t}$ denotes the state of the GNSS receiver at epoch $t$, $t \in (1, n)$, which involves the position ($\mathbf{p}_{r,t}$), velocity ($\mathbf{v}_{r,t}$) and receiver clock bias ($\delta_{r,t}$).

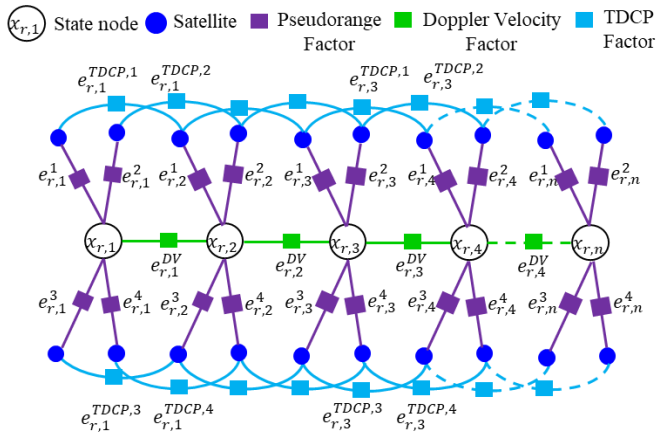

Fig. 2. Graph structure for a conventional pseudorange/Doppler/TDCP integration. The blue circle denotes the satellite. Only four satellites are drawn at each epoch for simplicity. The purple shaded rectangle denotes the pseudorange factor (e.g., $e_{r,t}^s$). The green shaded rectangle represents the Doppler velocity factor (e.g., $e_{r,t}^{DV}$). The light blue shaded rectangle denotes the TDCP factor. The white shaded circle stands for the state of the GNSS receiver.

The graph structure of the FGO for solving the GNSS positioning aided by the TDCP measurement is shown in Fig. 2. The subscript $n$ of state node denotes the total epochs of measurements considered in the FGO. Each state in the factor graph is connected using the Doppler velocity factor. Note that the formulation of the GNSS pseudorange/Doppler integration using the FGO is firstly presented in our latest work in [16]. However, we still present it here for the completeness of the derivation of the proposed integration in Section IV.

### A. Pseudorange Measurement Modeling

The pseudorange measurement from the GNSS receiver, $\rho_{r,t}^s$, is denoted as follows [27]:

$$\rho_{r,t}^s = r_{r,t}^s + c_L(\delta_{r,t} - \delta_{r,t}^s) + I_{r,t}^s + T_{r,t}^s + \varepsilon_{r,t}^s \qquad (3)$$

where $r_{r,t}^s$ is the geometric range between the satellite and the GNSS receiver. $c_L$ denotes the speed of light. $I_{r,t}^s$ represents the ionospheric delay; $T_{r,t}^s$ indicates the tropospheric delay. $\varepsilon_{r,t}^s$ represents the errors caused by the multipath effects, NLOS receptions, receiver noise, antenna phase-related errors. Meanwhile, the atmosphere effects ($T_{r,t}^s$ and $I_{r,t}^s$) are compensated using the conventional models (Saastamoinen and Klobuchar models, respectively) presented in RTKLIB [28].

The observation model for GNSS pseudorange measurement from a given satellite $s$ is represented as follows:

$$\rho_{r,t}^s = h_{r,t}^s(\mathbf{p}_{r,t}, \mathbf{p}_t^s, \delta_{r,t}) + \omega_{r,t}^s \qquad (4)$$

$$\text{with } h_{r,t}^s(\mathbf{p}_{r,t}, \mathbf{p}_t^s, \delta_{r,t}) = ||\mathbf{p}_t^s - \mathbf{p}_{r,t}|| + \delta_{r,t}$$

where the variable $\omega_{r,t}^s$ stands for the noise associated with the $\rho_{r,t}^s$. Therefore, we can get the error function ($\mathbf{e}_{r,t}^s$) for a given satellite measurement $\rho_{r,t}^s$ as follows:

$$||\mathbf{e}_{r,t}^s||_{\Sigma_{r,t}^s}^2 = ||\rho_{r,t}^s - h_{r,t}^s(\mathbf{p}_{r,t}, \mathbf{p}_t^s, \delta_{r,t})||_{\Sigma_{r,t}^s}^2 \qquad (5)$$

where $\Sigma_{r,t}^s$ denotes the covariance matrix. We calculate the $\Sigma_{r,t}^s$ based on the satellite elevation angle, signal, and noise ratio (SNR) following the work in [29].

### B. Doppler Measurements Modeling

Given the Doppler measurement ($d_{r,t}^1, d_{r,t}^2, ...$) of each satellite at an epoch $t$, the velocity ($\mathbf{v}_{r,t}$) of the GNSS receiver can be calculated using the weighted least square (WLS) method [30]. The detail of estimating the velocity of the GNSS receiver based on Doppler measurement can be found in our previous work in [16]. Therefore, the observation model for the velocity ($\mathbf{v}_{r,t}$) is expressed as follows:

$$\mathbf{v}_{r,t}^{DV} = h_{r,t}^{DV}(\mathbf{x}_{r,t+1}, \mathbf{x}_{r,t}) + \omega_{r,t}^{DV} \qquad (6)$$

$$\text{with } h_{r,t}^{DV}(\mathbf{x}_{r,t+1}, \mathbf{x}_{r,t}) = \begin{bmatrix} (p_{r,t+1,x} - p_{r,t,x})/\Delta t \\ (p_{r,t+1,y} - p_{r,t,y})/\Delta t \\ (p_{r,t+1,z} - p_{r,t,z})/\Delta t \end{bmatrix}$$

where the $\mathbf{v}_{r,t}^{DV}$ denotes the velocity measurements. The variable $\boldsymbol{\omega}_{r,t}^{DV}$ denotes the noise associated with the velocity measurements. The variable $\Delta t$ denotes the time difference between epoch $t$ and epoch $t + 1$. Therefore, we can get the error function ($\mathbf{e}_{r,t}^{DV}$) for a given Doppler velocity measurement $\mathbf{v}_{r,t}^{DV}$ as follows:

$$||\mathbf{e}_{r,t}^{DV}||_{\Sigma_{r,t}^{DV}}^2 = ||\mathbf{v}_{r,t}^{DV} - h_{r,t}^{DV}(\mathbf{x}_{r,t+1}, \mathbf{x}_{r,t})||_{\Sigma_{r,t}^{DV}}^2 \quad (7)$$

where $\Sigma_{r,t}^{DV}$ denotes the covariance matrix corresponding to the Doppler velocity measurement.

### C. TDCP Measurement Modeling

In terms of the measurements from the GNSS receiver, each carrier-phase measurement, $\psi_{r,t}^s$, is represented as follows [27].

$$\lambda \psi_{r,t}^s = r_{r,t}^s + c_L(\delta_{r,t} - \delta_{r,t}^s) + I_{r,t}^s + T_{r,t}^s + \lambda B_{r,t}^s + d\psi_{r,t}^s + \epsilon_{r,t}^s, \quad (8)$$

$$\text{where } B_{r,t}^s = \psi_{r,0,t} - \psi_{0,t}^s + N_{r,t}^s$$

where $B_{r,t}^s$ is the carrier-phase bias. The variable $\lambda$ denotes the carrier wavelength. The variable $d\psi_{r,t}^s$ denotes the carrier-phase correction term including antenna phase offsets and variations, station displacement by earth tides, phase windup effect, and relativity correction on the satellite clock. The detailed formulation of the carrier-phase correction can be found in [28]. $\epsilon_{r,t}^s$ represents the errors caused by the multipath effects, NLOS receptions, receiver noise, antenna delay. The variable $\psi_{r,0,t}$ represents the initial phase of the receiver local oscillator. Similarly, the $\psi_{0,t}^s$ stands for the initial phase of the transmitted navigation signal from the satellite. The variable $N_{r,t}^s$ denotes the carrier-phase integer ambiguity. Typically, the $N_{r,t}^s$ is an unknown variable to be estimated during the utilization of the carrier-phase measurements. By differencing the raw carrier-phase measurements between two epochs, the TDCP measurement ($\Delta \lambda \psi_{r,t,t+1}$) can be derived as follows:

$$\Delta \lambda \psi_{r,t,t+1} = (r_{r,t+1}^s - r_{r,t}^s) + c_L(\delta_{r,t+1} - \delta_{r,t}) + (\epsilon_{r,t+1}^s - \epsilon_{r,t}^s) \quad (9)$$

where the unknown integer ambiguity term is eliminated by the differencing operation. Therefore, the observation model for the TDCP measurement from a given satellite $s$ is represented as follows:

$$\Delta \lambda \psi_{r,t,t+1}^s = h_{r,t}^{TDCP,s}(\mathbf{p}_{r,t}, \mathbf{p}_t^s, \delta_{r,t}, \mathbf{p}_{r,t+1}, \mathbf{p}_{t+1}^s, \delta_{r,t+1}) + \omega_{r,t}^{TDCP,s} \quad (10)$$

with $h_{r,t}^{TDCP,s}(\mathbf{p}_{r,t}, \mathbf{p}_t^s, \delta_{r,t}, \mathbf{p}_{r,t+1}, \mathbf{p}_{t+1}^s, \delta_{r,t+1}) = (||\mathbf{p}_t^s - \mathbf{p}_{r,t}|| + \delta_{r,t}) - (||\mathbf{p}_{t+1}^s - \mathbf{p}_{r,t+1}|| + \delta_{r,t})$

where the variable $\omega_{r,t}^{TDCP,s}$ stands for the noise associated with the TDCP measurement. Therefore, we can get the error function ($\mathbf{e}_{r,t}^{TDCP,s}$) for the given TDCP measurements ($\psi_{r,t}^s, \psi_{r,t+1}^s$) as follows:

$$||\mathbf{e}_{r,t}^{TDCP,s}||_{\Sigma_{r,t}^{TDCP,s}}^2 = ||\Delta \lambda \psi_{r,t,t+1}^s - h_{r,t}^{TDCP,s}(\mathbf{p}_{r,t}, \mathbf{p}_t^s, \delta_{r,t}, \mathbf{p}_{r,t+1}, \mathbf{p}_{t+1}^s, \delta_{r,t+1})||_{\Sigma_{r,t}^{TDCP,s}}^2 \quad (11)$$

where $\Sigma_{r,t}^{TDCP,s}$ denotes the covariance matrix. Similarly, we calculate the $\Sigma_{r,t}^{TDCP,s}$ based on the satellite elevation angle, signal, and noise ratio (SNR) following the work in [29].

### D. Pseudorange/Doppler/TDCP Fusion Via FGO

Based on the factors derived above, the combined objective function can be formulated as follows:

$$\boldsymbol{\chi}^* = \arg\min_{\boldsymbol{\chi}} \sum_{s,t} (||\mathbf{e}_{r,t}^{DV}||_{\Sigma_{r,t}^{DV}}^2 + ||\mathbf{e}_{r,t}^s||_{\Sigma_{r,t}^s}^2 + ||\mathbf{e}_{r,t}^{TDCP,s}||_{\Sigma_{r,t}^{TDCP,s}}^2) \quad (12)$$

The variable $\boldsymbol{\chi}^*$ denotes the optimal estimation of the state sets, which can be estimated by solving the objective function above iteratively.

## IV. GNSS POSITIONING AIDED BY WINDOW CARRIER-PHASE

This section presents the methodology of the GNSS positioning aided by the proposed WCP constraint. The graph structure of the FGO for solving the GNSS positioning aided by the WCP constraint is shown in Fig.3 where the TDCP in Fig. 2 is replaced by the proposed WCP.

### A. WCP Constraint Modeling

Fig. 3 shows that four satellites (satellites 1, 2, 3, and 4) are received repetitively from epoch 1 to epoch $n$ which are involved in the factor graph. Taking satellite 1 as an example, the continuous observed carrier-phase measurements can be stacked into the following form:

$$\begin{bmatrix} \lambda\psi_{r,1}^s \\ \lambda\psi_{r,2}^s \\ ... \\ \lambda\psi_{r,N_k^s}^s \end{bmatrix} = \begin{bmatrix} r_{r,1}^s + m_{r,1}^s \\ r_{r,2}^s + m_{r,2}^s \\ ... \\ r_{r,N_k^s}^s + m_{r,N_k^s}^s \end{bmatrix} + \begin{bmatrix} \lambda B_{r,1}^s \\ \lambda B_{r,2}^s \\ ... \\ \lambda B_{r,N_k^s}^s \end{bmatrix} + \begin{bmatrix} \epsilon_{r,1}^s \\ \epsilon_{r,2}^s \\ ... \\ \epsilon_{r,N_k^s}^s \end{bmatrix} \quad (13)$$

With $m_{r,t}^s = c_L(\delta_{r,t} - \delta_{r,t}^s) + I_{r,t}^s + T_{r,t}^s + d\psi_{r,t}^s$

where the $N_k^s$ indicates that satellite 1 is tracked by the GNSS receiver continuously for $N_k^s$ epochs. The carrier-phase measurements involved in (13) form a window carrier-phase of satellite 1 with a window size of $N_k^s$. The variable $k$ denotes the index of the window carrier-phase constraint inside the factor graph. The $m_{r,t}^s$ is defined for simplicity.

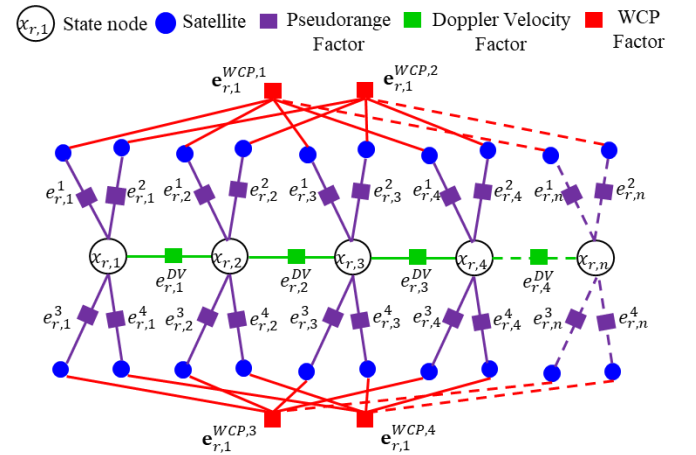

Fig. 3. Graph structure for the proposed pseudorange/Doppler/WCP integration. The blue circles denote the satellite. The purple shaded rectangle denotes the pseudorange factor (e.g $e_{r,t}^s$). The green shaded rectangle represents the Doppler velocity factor (e.g $e_{r,t}^{DV}$). The red shaded rectangle denotes the WCP factor. The white shaded circle stands for the state of the GNSS receiver.

Giving the fact that the carrier-phase measurements inside the window share the same integer ambiguity, the (13) can be rewritten as follows:

$$\begin{bmatrix} \lambda\psi_{r,1}^1 \\ \lambda\psi_{r,2}^1 \\ ... \\ \lambda\psi_{r,N_k^s}^1 \end{bmatrix} = \begin{bmatrix} r_{r,t}^s + m_{r,1}^s \\ r_{r,2}^s + m_{r,2}^s \\ ... \\ r_{r,N_k^s}^s + m_{r,N_k^s}^s \end{bmatrix} + \lambda B_{r,N_k^s}^s \begin{bmatrix} 1 \\ 1 \\ 1 \end{bmatrix} + \begin{bmatrix} \epsilon_{r,1}^s \\ \epsilon_{r,2}^s \\ ... \\ \epsilon_{r,N_k^s}^s \end{bmatrix} \quad (14)$$

where the $B_{r,N_k^s}^s$ denotes the shared integer ambiguity value for satellite 1 within the continuous $N_k^s$ epochs. The compact form of the equation (14) can be further organized by multiply a left null space matrix $\mathbf{G}_{r,k}^s$ on both sides of equation (14) and the following form is obtained:

$$\mathbf{G}_{r,k}^s \begin{bmatrix} \lambda\psi_{r,1}^1 \\ \lambda\psi_{r,2}^1 \\ ... \\ \lambda\psi_{r,N_k^s}^1 \end{bmatrix} = \mathbf{G}_{r,k}^s \begin{bmatrix} r_{r,1}^s + m_{r,1}^s \\ r_{r,2}^s + m_{r,2}^s \\ ... \\ r_{r,N_k^s}^s + m_{r,N_k^s}^s \end{bmatrix} + \mathbf{G}_{r,k}^s \begin{bmatrix} \epsilon_{r,1}^s \\ \epsilon_{r,2}^s \\ ... \\ \epsilon_{r,N_k^s}^s \end{bmatrix} \quad (15)$$

$$\text{with } \mathbf{G}_{r,k}^s \begin{bmatrix} 1 \\ 1 \\ ... \\ 1 \end{bmatrix} = \begin{bmatrix} 0 \\ 0 \\ ... \\ 0 \end{bmatrix}$$

Therefore, the derived (15) is free of ambiguity variable which is eliminated by a left multiplication of the left null space matrix $\mathbf{G}_{r,k}^s$. The construction of the $\mathbf{G}_{r,k}^s$ can be found in Appendix A of this paper.

Therefore, the observation model for WCP constraint from a given satellite $s$ being tracked by $N_k^s$ epochs are represented as follows:

$$\mathbf{G}_{r,k}^s \begin{bmatrix} \lambda\psi_{r,1}^1 \\ \lambda\psi_{r,2}^1 \\ ... \\ \lambda\psi_{r,N_k^s}^1 \end{bmatrix} = \mathbf{G}_{r,k}^s \begin{bmatrix} h_{r,1}^{WCP,s}(\mathbf{p}_{r,1}, \mathbf{p}_1^s, \delta_1) \\ h_{r,2}^{WCP,s}(\mathbf{p}_{r,2}, \mathbf{p}_2^s, \delta_2) \\ ... \\ h_{r,N_k^s}^{WCP,s}(\mathbf{p}_{r,N_k^s}, \mathbf{p}_{N_k^s}^s, \delta_{N_k^s}) \end{bmatrix} + \mathbf{G}_{r,k}^s \begin{bmatrix} \epsilon_{r,1}^s \\ \epsilon_{r,2}^s \\ ... \\ \epsilon_{r,N_k^s}^s \end{bmatrix} \quad (16)$$

$$\text{with } h_{r,N_k^s}^{WCP,s}\left(\mathbf{p}_{r,N_k^s}, \mathbf{p}_{N_k^s}^s, \delta_{1,N_k^s}\right) = ||\mathbf{p}_{N_k^s}^s - \mathbf{p}_{r,N_k^s}|| + \delta_{r,N_k^s}$$

Therefore, we can get the error function ($\mathbf{e}_{r,k}^{WCP,s}$) for a WCP constraint as follows:

$$||\mathbf{e}_{r,k}^{WCP,s}||_{\Sigma_{r,k}^{WCP,s}}^2 = \left\| \mathbf{G}_{r,k}^s \begin{bmatrix} \lambda\psi_{r,1}^1 \\ \lambda\psi_{r,2}^1 \\ ... \\ \lambda\psi_{r,N_k^s}^1 \end{bmatrix} - \right.$$

$$\left. \mathbf{G}_{r,k}^s \begin{bmatrix} h_{r,1}^{WCP,s}(\mathbf{p}_{r,1}, \mathbf{p}_1^s, \delta_1) \\ h_{r,2}^{WCP,s}(\mathbf{p}_{r,2}, \mathbf{p}_2^s, \delta_2) \\ ... \\ h_{r,N_k^s}^{WCP,s}\left(\mathbf{p}_{r,N_k^s}, \mathbf{p}_{N_k^s}^s, \delta_{N_k^s}\right) \end{bmatrix} \right\|_{\Sigma_{r,k}^{WCP,s}}^2 \quad (17)$$

where $\Sigma_{r,k}^{WCP,s}$ denotes the covariance matrix associated with the WCP constraint which can be calculated as follows:

$$\Sigma_{r,k}^{WCP,s} = \mathbf{G}_{r,k}^s \begin{bmatrix} \epsilon_{r,1}^s & 0 & 0 & 0 \\ 0 & \epsilon_{r,2}^s & 0 & 0 \\ 0 & 0 & ... & 0 \\ 0 & 0 & 0 & \epsilon_{r,N_k^s}^s \end{bmatrix} \mathbf{G}_{r,k}^s{}^{\mathrm{T}} \quad (18)$$

The WCP factor ($\mathbf{e}_{r,k}^{WCP,s}$) is denoted by the red shaded rectangles in Fig. 3 which connects the same satellite received in multiple epochs across the factor graph.

### B. Comparison Between the Conventional TDCP and the Proposed WCP

Different from the conventional TDCP constraint which only explores the kinematic correlation between the two consecutive epochs, the proposed WCP constraint effectively makes use of the correlation of shared integer ambiguity within multiple epochs inside the window. Meanwhile, the involved covariance matrix ($\Sigma_{r,k}^{WCP,s}$) effectively maintained the cross-correlation using the multiplication of $\mathbf{G}_{r,k}^s$. Specifically, a similar format of the formulation can also be derived for TDCP measurements given a set of continuously tracked GNSS measurements from the same satellite as follows:

$$\begin{bmatrix} \Delta\lambda\psi_{r,1}^1 \\ \Delta\lambda\psi_{r,2}^1 \\ ... \\ \Delta\lambda\psi_{r,N_k^s}^1 \end{bmatrix} = \mathbf{D}_{r,k}^s \begin{bmatrix} r_{r,t}^s + m_{r,1}^s \\ r_{r,2}^s + m_{r,2}^s \\ ... \\ r_{r,N_k^s}^s + m_{r,N_k^s}^s \end{bmatrix} + \mathbf{D}_{r,k}^s \begin{bmatrix} \lambda B_{r,1}^s \\ \lambda B_{r,2}^s \\ ... \\ \lambda B_{r,N_k^s}^s \end{bmatrix} +$$

$$\mathbf{D}_{r,k}^s \begin{bmatrix} \epsilon_{r,1}^s \\ \epsilon_{r,2}^s \\ ... \\ \epsilon_{r,N_k^s}^s \end{bmatrix} \quad (19)$$

where the $\Delta\lambda\psi_{r,1}^1$ denotes the TDCP measurements of satellite 1. The $\mathbf{D}_{r,k}^s$ denotes the time difference matrix which is employed to remove the ambiguity term in TDCP operation which can be denoted as follows (5 continuously tracked GNSS measurements as an example):

$$\mathbf{D}_{r,k}^s = \begin{bmatrix} -1 & 1 & 0 & 0 & 0 \\ 0 & -1 & 1 & 0 & 0 \\ 0 & 0 & -1 & 1 & 0 \\ 0 & 0 & 0 & -1 & 1 \\ 0 & 0 & 0 & 0 & 0 \end{bmatrix} \quad (20)$$

The $\mathbf{D}_{r,k}^s$ matrix is sparse and only enables the time difference between consecutive epochs which is the major difference between the TDCP and the proposed WCP. The analysis of the matrix $\mathbf{G}_{r,k}^s$ which is a densely correlated matrix, and the $\mathbf{D}_{r,k}^s$ with real data is to be presented in Section V.

### C. Cycle Slip Accommodation and Selection of the Window Size of the WCP

**Cycle Slip Accommodation**: In typical urban canyons, the occasional GNSS signal blockage and reflection can cause the cycle slip problem which can therefore violate the assumption of shared integer ambiguity inside a WCP. From the measurement domain, the cycle slip phenomenon can cause significant inconsistency in the consecutive carrier-phase measurements. Thanks to the factor graph optimization which involves a large number of redundant measurements, this paper employs the robust M-estimator [26] to detect and mitigate the effects of the cycle slip. Our previous work [31] shows that the M-estimator can effectively mitigate the unexpected visual outlier measurements in a visual/inertial integrated positioning system. A more detailed discussion about the cycle slip mitigation analysis using M-estimator can be found in Section V.

*Window size selection of the WCP*: Theoretically, the optimal window size ($N_k^S$) can be selected by maximizing the exploration of the time correlation of the states involved. Therefore, it is good to select the largest window size as long as the assumption of constant integer ambiguity inside the WCP holds. However, the invariant ambiguity inside the WCP for a long period is hard to be satisfied in urban canyons due to the potential cycle slip. Therefore, we set a threshold for the $N_k^S$ denoted by the $N_{k,MAX}^S$. After the size of WCP exceeds the $N_{k,MAX}^S$, a WCP should be constructed.

### D. Pseudorange/Doppler/WCP Fusion Via FGO

Based on the factors derived above, the combined objective function proposed in this paper can be formulated as follows:

$$\boldsymbol{\chi}^* = \arg\min_{\boldsymbol{\chi}} \sum_{s,t,k} (||\mathbf{e}_{r,t}^{DV}||_{\Sigma_{r,t}^{DV}}^2 + ||\mathbf{e}_{r,t}^S||_{\Sigma_{r,t}^S}^2 + ||\rho(\mathbf{e}_{r,k}^{WCP,S})||_{\Sigma_{r,k}^{WCP,S}}^2) \quad (21)$$

The variable $\boldsymbol{\chi}^*$ denotes the optimal estimation of the state sets, which can be estimated by solving the objective function above iteratively. The operator $\rho(*)$ denotes the robust function of the M-estimator [26] and the Cauchy function is applied as can be formulated as follows:

$$\rho(\mathbf{e}_{r,k}^{WCP,S}) = \frac{k_\rho^2}{2}\log\left(1 + \frac{(\mathbf{e}_{r,k}^{WCP,S})^2}{k_\rho^2}\right) \quad (22)$$

where the $k_\rho$ denotes the kernel value (parameter) which dominates the non-convexity of the Cauchy function. Meanwhile, the performance of the Cauchy function relies heavily on the selection of the $k_\rho$ and the analysis is to be presented in Section V. Both the objective equations (12) and (21) are solved using the Levenberg-Marquardt algorithm via the state-of-the-art Ceres-solver [32].

## V. Experiment Results and Discussion

### A. Experiment Setup

*Experimental scenes*: To verify the effectiveness of the proposed method, two experiments were conducted in typical urban canyons in Hong Kong. Fig. 4 shows the data collection vehicle (Fig. 4-(a)), scenes in urban canyon 1 (Fig. 4-(b)) and 2 (Fig. 4-(c)), respectively. Both of the urban scenarios contain static buildings, trees, and dynamic objects, such as double-decker buses, which can lead to numerous GNSS outlier measurements.

*Sensor setups*: Fig. 4-(a) shows the data collection vehicle and the detail of the data collection vehicle can be found through our open-sourced UrbanNav [33] dataset [1]. In both experiments, a u-blox M8T GNSS receiver was used to collect raw GNSS measurements at a frequency of 1 Hz. Besides, the

NovAtel SPAN-CPT, a GNSS (GPS, GLONASS, and Beidou) RTK/INS (fiber-optic gyroscopes, FOG) integrated navigation system was used to provide ground truth of positioning. The gyro bias in-run stability of the FOG is 1 degree per hour, and its random walk is 0.067 degrees per hour. The baseline between the rover and the GNSS base station is about 7 km. Meanwhile, we post-process the data from SPAN-CPT using the inertial explorer software from NovAtel to guarantee the accuracy of the ground truth of positioning. All the data were collected and synchronized using a robot operation system (ROS) [34]. The coordinate systems between all the sensors were calibrated before the experiments.

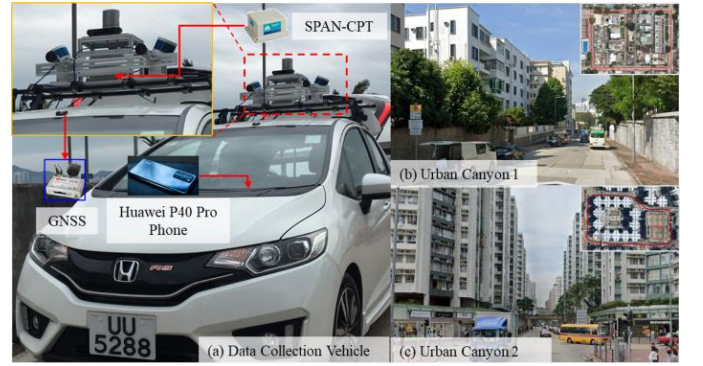

Fig. 4. Demonstration of the evaluated urban canyons 1 and 2. The red curves inside the Figures denote the trajectories of the tests.

*Evaluation metrics*: We analyze the performance of positioning by comparing five different methods, as shown below to validate the effectiveness of the proposed method in improving the GNSS positioning.

(a) **u-blox** [35]: the commercial GNSS positioning solution from the u-blox M8T receiver.

(b) **WLS** [28]: weighted least squares positioning.

(c) **PSR-DOP (Sol 1)**: the pseudorange and Doppler integration using FGO based on [16].

(d) **PSR-DOP-TDCP (Sol 2)**: the pseudorange, Doppler, and TDCP measurements integration using FGO (corresponding to the factor graph in Fig. 2).

(e) **PSR-DOP-WCP (Sol 3)**: the pseudorange, Doppler, and WCP measurements integration using FGO (corresponding to the factor graph in Fig. 3) with maximum window size ($N_{k,MAX}^S$) of 6.

### B. Experimental Evaluation in Urban Canyon 1

#### 1) Positioning Performance Evaluation

The positioning results of the listed five methods are shown in Table 1. The first column shows the 2D positioning error of the u-blox receiver. The positioning result is based on standard NMEA [36] messages from the u-blox receiver. A mean error of 6.23 meters was obtained, with a standard deviation of 7.31 meters. The maximum error reached 38.53 meters. The second column shows the positioning result using the raw pseudorange measurements from the u-blox receiver and positioning based on WLS. Similarly, the weighting scheme was taken from [29] and is based on the satellite elevation angle and the signal-to-noise ratio (SNR). The positioning error decreased to 3.10 meters with a standard deviation of 1.95 meters. The maximum error also decreased to about 11 meters. Fig. 5 shows the positioning error throughout the evaluation where the red and green denote the solutions from u-blox and WLS, respectively. The positioning error fluctuate occasionally during the test due to the unexpected GNSS outlier measurements arising from the blockage and reflections by surrounding buildings.

With the help of the integration of the Doppler frequency measurements, the positioning error decreases to 2.14 meters with the standard deviation decreasing to 1.01 meters. The improved results show that the Doppler frequency measurement can help to mitigate the effects of the GNSS outlier measurements which can also be seen from Fig. 5 with a smoother error curve by the method of PSR-DOP. However, the jump can still be seen (e.g. epoch A annotated by the circle in Fig. 5) as the Doppler measurement can also be biased by the multipath effects [37]. After applying the carrier-phase measurements using the TDCP technique (corresponding to the factor graph in Fig. 2), the positioning error decreases slightly to 2.01 meters (PSR-DOP-TDCP) with a maximum error of 5.69 meters. However, a more significant improvement can be seen in the standard deviation which means that smoother results are obtained. The results show that the high-accuracy carrier-phase measurements can help to smooth the positioning against GNSS outlier measurements.

With the help of the proposed WCP constraint (corresponding to the factor graph in Fig. 3), the positioning error decreases to only 1.76 meters which is almost a lane-level accuracy which shows that the effectiveness of the proposed method. Moreover, the standard deviation decreases to only 0.57 meters with a maximum error of 4.06 meters. The significantly decreased standard deviation indicates that the proposed WCP constraint can effectively constrain the multiple states more tightly compared with the TDCP which only involves the two pairwise states. Fig. 6 shows the trajectories on the evaluated five methods. Epoch B in Fig. 5 is corresponding to Fig. 6-(a) located in an intersection with tall buildings. As a result, the positioning error of the WLS is larger than 10 meters. With the help of the WCP, the error decreases from 5.6 meters for PSR-DOP-TDCP to only 4.06 meters. Fig. 6-(b) shows a scene with dense foliage which can cause GNSS signal blockage and diffractions. As a result, the solution from WLS deviates significantly from the ground truth. However, smooth results are still achieved with the proposed method (PSR-DOP-WCP).

In short, both the Doppler frequency and the carrier-phase measurements can help to improve the positioning accuracy by mitigating the impacts of the GNSS outlier measurements. We gradually compare the improvement from both measurements. The utilization of the TDCP (PSR-DOP-TDCP) obtains similar accuracy compared with the PSR-DOP although with a slight decrease of STD in the evaluated dataset. Fortunately, the proposed method (PSR-DOP-WCP) pushes the positioning performance to a higher accuracy with a mean error of 1.76 meters and an STD of only 0.57 meters.

Table 1. Performance of the listed five methods in urban canyon 1 (Max: maximum error; STD: standard deviation; MEAN: mean error; Sol 1: PSR-DOP, Sol 2: PSR-DOP-TDCP, Sol 3: PSR-DOP-WCP)

| Item (m) | u-blox | WLS | Sol 1 | Sol 2 | Sol 3 |
|---|---|---|---|---|---|
| **MEAN** | 6.23 | 3.10 | 2.14 | 2.01 | 1.76 |
| **STD** | 7.31 | 1.95 | 1.01 | 0.82 | 0.57 |
| **Max** | 38.53 | 11.23 | 6.52 | 5.69 | 4.06 |

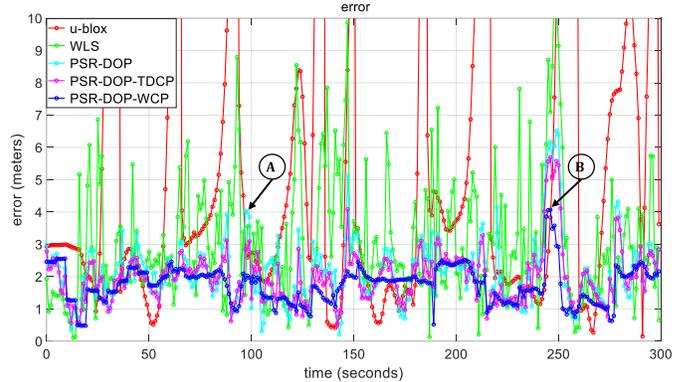

Fig. 5. 2D positioning errors of the evaluated five methods in urban canyon 1. The red curve denotes the 2D error from the commercial solution of the u-blox receiver. The green, cyan, magenta, and blue colors denote the WLS, PSR-DOP, PSR-DOP-TDCP, and PSR-DOP-WCP, respectively.

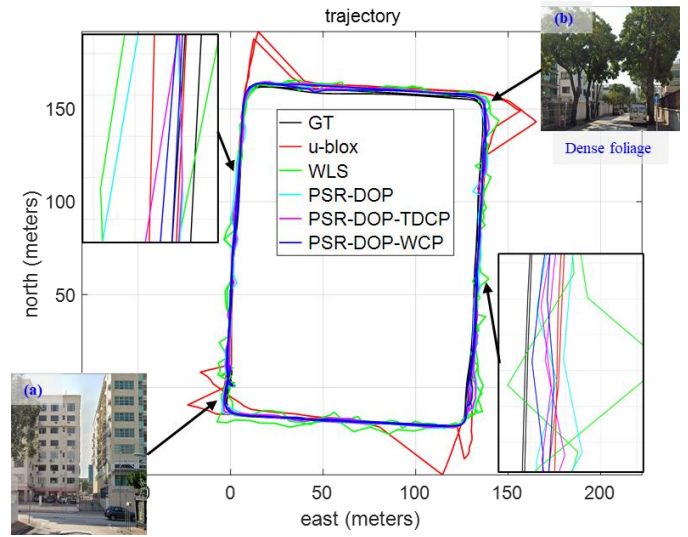

Fig. 6. 2D positioning trajectories of the evaluated five methods in urban canyon 1. The black and red curves denote the 2D trajectories of the ground truth and the commercial solution of the u-blox receiver. The green, cyan, magenta, and blue colors denote the WLS, PSR-DOP, PSR-DOP-TDCP, and PSR-DOP-WCP, respectively.

### 2) Analysis of the Impact of Maximum Window Size

Since the proposed WCP constraint is highly related to the maximum window size ($N_{k,MAX}^s$). Table 2 shows how the $N_{k,MAX}^s$ impacts the performance of the overall integration (PSR-DOP-WCP). Therefore, we present the accuracy of the PSR-DOP-WCP under the different values of $N_{k,MAX}^s$. The maximum window size of 6 is corresponding to the evaluation in Section B-(1). Interestingly, after slightly increase the $N_{k,MAX}^s$, both the mean and standard deviation decrease with the mean error decreasing from 1.76 to 1.69 meters. This is due to fact that the larger $N_{k,MAX}^s$ enables the stronger correlation between the states. However, after almost doubling $N_{k,MAX}^s$, the mean error slightly increases to 1.82 meters. A similar phenomenon is also seen by increasing the $N_{k,MAX}^s$ to 25 and 35, respectively. This is caused by the violation of shared ambiguity variables inside the WCP due to the cycle slip. Although the proposed method in this paper employs the M-estimator to mitigate the effects of the cycle slip, the error caused still exists.

Fig. 7 shows the details of the errors under different values of $N_{k,MAX}^s$. The representative epoch is epoch A (denoted by a circle in Fig. 7) where the green curve ($N_{k,MAX}^s = 9$) obtains better accuracy than all other methods. In addition to $N_{k,MAX}^s$, the maximum size of the WCP can also be determined by the times that one satellite is being tracked. The histograms (Fig. 7-(a) and Fig. 7-(b)) denote the distributions of the window size of the WCP during two tests. It is shown that the majority of the WCP can have a size of the $N_{k,MAX}^s$. However, the $N_k^s$ of some satellites is even smaller than 10 due to lost track occasionally. In short, the performance of the WCP constraint is highly related to the window size, and too large a size does not necessarily lead to better results due to the existence of occasional cycle slips.

Table 2. Performance of the proposed method under different maximum window sizes ($N_{k,MAX}^s$) assumptions.

| $N_{k,MAX}^s$ | 6 | 9 | 16 | 25 | 35 |
|---|---|---|---|---|---|
| **MEAN (m)** | 1.76 | 1.69 | 1.82 | 1.89 | 1.93 |
| **STD (m)** | 0.57 | 0.48 | 0.62 | 0.72 | 0.82 |
| **Max (m)** | 4.06 | 3.16 | 4.67 | 5.43 | 6.01 |

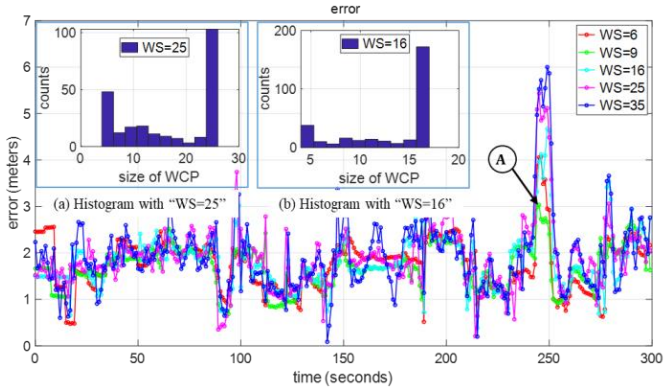

Fig. 7. 2D positioning errors of the proposed method under the different values of $N_{k,MAX}^s$. The "WS" in the legend denotes the $N_{k,MAX}^s$. The x-axis denotes time and the y-axis represents the errors.

### C. Experimental Evaluation in Urban Canyon 2

#### 1) Positioning Performance Evaluation

To challenge the performance of the proposed method, another experiment was conducted in a denser urban canyon 2 with significantly higher building structures. Table 3 shows that a mean error of 31.02 meters is achieved via the commercial solution from the u-blox receiver. Similar to the experimental results in urban canyon 1, improved accuracy is obtained by the WLS with the mean error decreasing to 16.66 meters. Fortunately, the integration of the pseudorange and Doppler measurements using FGO (PSR-DOP) significantly reduce the error to only 3.9 meters with a significantly decreased standard deviation. Interestingly, the utilization of the TDCP constraint even degrades the overall performance with the mean error increasing to 5.21 meters. This is due to the significantly increased percentage of the cycle slip phenomenon in urban canyons 2 with denser building structures. As a result, the TDCP becomes an incorrect constraint due to the violation of the constant integer ambiguity between consecutive epochs of measurements. However, after applying the proposed WCP, the mean error decreases to only 2.96 meters with the standard deviation decreasing to only 1.89 meters. The detailed 2D positioning errors of the evaluated five methods can also be shown in Fig. 8. Different from the TDCP, the WCP involves multiple carrier-phase measurements and the impacts of the single-cycle slip can be further resisted by the measurements redundancy of window constraint. This significant improvement again shows the effectiveness of the proposed method in improving the GNSS positioning even in such a dense urban canyon.

Table 3. Performance of the listed five methods in urban canyon 2 (Max: maximum error; STD: standard deviation; MEAN: mean error; Sol 1: PSR-DOP, Sol 2: PSR-DOP-TDCP, Sol 3: PSR-DOP-WCP)

| Item (m) | u-blox | WLS | Sol 1 | Sol 2 | Sol 3 |
|---|---|---|---|---|---|
| **MEAN** | 31.02 | 16.66 | 3.90 | 5.21 | 2.96 |
| **STD** | 37.69 | 12.82 | 2.29 | 2.38 | 1.89 |
| **Max** | 177.6 | 56.23 | 12.84 | 12.52 | 12.02 |

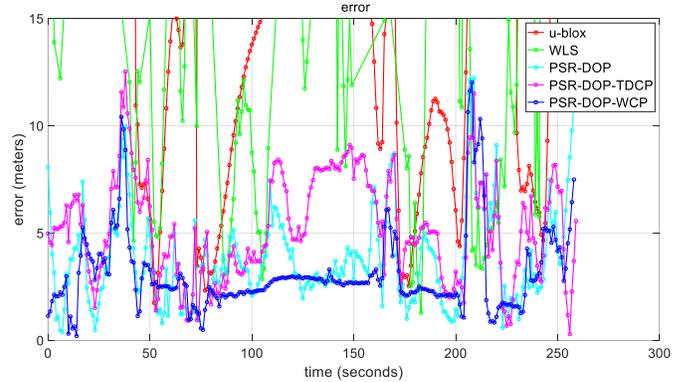

Fig. 8. 2D positioning errors of the evaluated five methods in urban canyon 2. The red curve denotes the 2D error from the commercial solution of the u-blox receiver. The green, cyan, magenta, and blue colors denote the WLS, PSR-DOP, PSR-DOP-TDCP, and PSR-DOP-WCP, respectively.

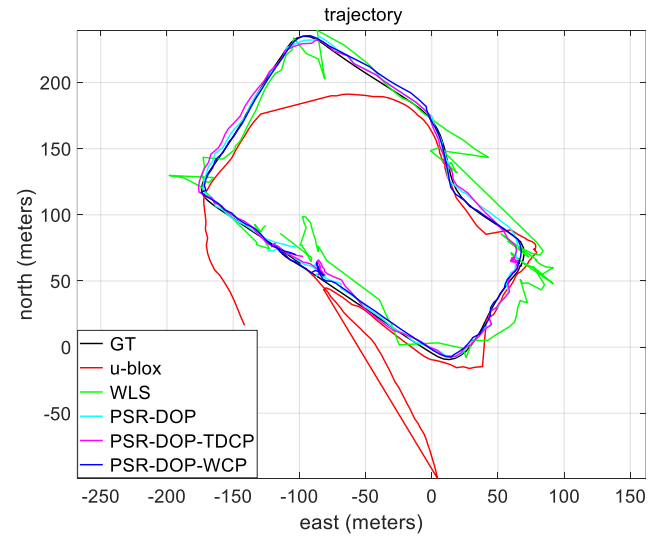

Fig. 9. 2D positioning trajectories of the evaluated five methods in urban canyon 2. The black and red curves denote the 2D trajectories of the ground truth and the commercial solution of the u-blox receiver. The green, cyan, magenta, and blue colors denote the WLS, PSR-DOP, PSR-DOP-TDCP, and PSR-DOP-WCP, respectively.

| Item (m) | H (1) | H (2) | C (1) | C (2) | C (4) |
|----------|-------|-------|-------|-------|-------|
| **MEAN** | Diverge | Diverge | 3.62 | 3.11 | 3.05 |
| **STD** | Diverge | Diverge | 2.04 | 1.78 | 1.68 |
| **Max** | Diverge | Diverge | 13.11 | 11.20 | 10.76 |

Fig. 9 shows the trajectories of the evaluated five methods. Both the solutions from the u-blox (red curve) and the WLS (green curve) deviate significantly from the ground truth (black curve) due to the GNSS outlier measurements. With the help of the proposed WCP constraint, a significantly smoother trajectory (blue curve) is obtained.

### 2) Analysis of the Cycle Slip Accommodation via M-estimator

As mentioned in Section IV-C, the cycle slip is accommodated by the M-estimator by mitigating the effects of the inconsistency in integer ambiguity. This is based on the assumption that the cycle slip can induce inconsistency in integer ambiguity where large residuals (see (17)) can be observed after solving the (21). However, the performance of the M-estimator relies heavily on the selection of its kernel parameter. To see how the selection of the kernel parameter impacts the performance of the proposed WCP with different kinds of M-estimators, Table 4 presents the positioning performance of the proposed PSR-DOP-WCP under different M-estimator setups. The Huber-based M-estimator (see Appendix B regarding the formulation of the Huber function) fails to resist the inconsistency of integer ambiguity caused by the cycle slip for both selected kernel parameters where the PSR-DOP-WCP diverges. This is mainly due to the convex surface property of the Huber-based M-estimator. As a result, the Huber cannot mitigate the impacts of the gross outlier. The Huber and Cauchy functions with different kernel values ($k_\rho$) are shown in Fig. 10. The red shaded area denotes the potential outlier area with larger residuals. The light green area denotes the typical healthy measurements area with smaller residuals. The cyan curve denotes the conventional squared loss function where the gross outlier of the WCP constraint can cause large loss, leading to a large final position estimation error. The Huber-based robust function can mitigate the impacts of the outliers (e.g., red curve), compared with the conventional squared loss function. However, the outlier with a large residual can still mislead the final position estimation. For example, a residual of 50 can cause a loss of about 100 using the Huber (2) which can be seen by the red circle of the marker in Fig. 10. Differently, the Cauchy function is a cut-off function that can significantly mitigate the effects of the gross outlier. For example, a residual of 50 can cause a loss of only 4 using the Cauchy (1) which can be seen by the blue asterisk of the marker in Fig. 10. As a result, a mean error of 3.62 meters is obtained using Cauchy based M-estimator with a kernel value of 1. The mean error decreases to 3.11 meters with a kernel value of 2. A similar mean error is obtained after increasing the kernel value to 4 which is an appropriate kernel to accommodate the cycle slip phenomenon.

Table 4. Positioning error of the PSR-DOP-WCP under different M-estimator setups in urban canyon 2 (H (1): Huber function with kernel value ($k_\rho$) of 1; C (1): Cauchy function with kernel ($k_\rho$) value of 1)

To show the detail that how the M-estimator helps to mitigate the impacts of the cycle slip phenomenon, as Fig. 11 shows that we present the residuals of the WCP constraints in urban canyon 2, using Huber and Cauchy functions, respectively. With the help of the Cauchy function, the gross residuals are mitigated with all the residuals ranging between $[-4, 4]$. Conversely, the residuals involve long-tail with large values using the Huber function.

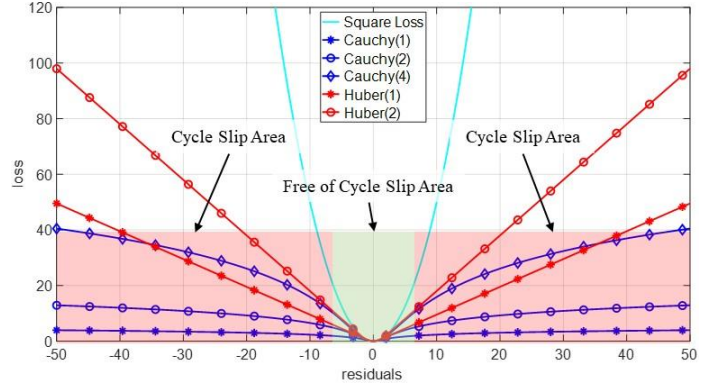

Fig. 10. Comparisons of the robust functions with different kernel values ($k_\rho$).

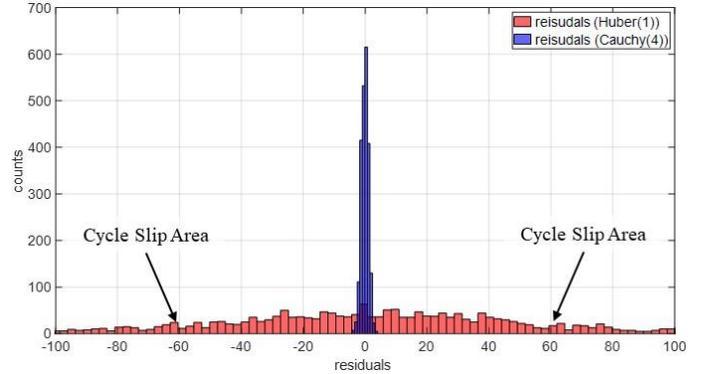

Fig. 11. Histograms of residuals of the WCP constraints using the Cauchy and Huber functions. The x-axis denotes the residuals calculated using the (17) and the y-axis represents the counts.

### 3) Analysis of the TDCP and Left Null Space Matrices

As shown in the experimental results of the urban canyons 1 and 2, the improved positioning results are obtained using the proposed WCP constraint based on the $\mathbf{G}_{r,k}^s$, compared with the conventional TDCP constraint based on the $\mathbf{D}_{r,k}^s$ (see (20)). It is interesting to compare the difference between the two matrices with real data. To show the detail of the left null space matrix $\mathbf{G}_{r,k}^s$ of WCP constraint and the TDCP matrix $\mathbf{D}_{r,k}^s$, we select a window for a satellite being observed for 11 epochs in the urban canyon 2. The $\mathbf{G}_{r,k}^s$ of WCP constraint is shown on the right side of Fig. 12. Similarly, the $\mathbf{D}_{r,k}^s$ for the TDCP measurements, is shown on the left side of Fig. 12. We can see that the off-diagonal components of the $\mathbf{D}_{r,k}^s$ for the TDCP are zero which neglects the cross-correlation between the multiple epochs of carrier-phase measurements. According to our previous work in [16], the GNSS measurements are highly environmentally dependent and time-correlated in urban scenarios. Differently, the $\mathbf{G}_{r,k}^s$ for the proposed WCP is a dense matrix that effectively explores the correlation between measurements between consecutive epochs which is the major difference between the proposed WCP and the conventional TDCP. As a result, the WCP constraint significantly increases the robustness of the estimator against outlier measurements and leads to smoother position estimation with the help of multiple epoch constraints.

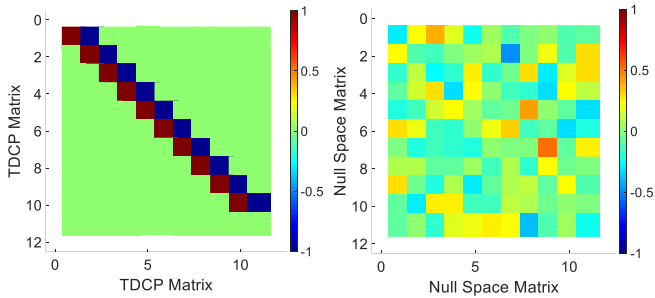

Fig. 12. Comparison of the $\mathbf{D}_{r,k}^s$ for TDCP constraint and the $\mathbf{G}_{r,k}^s$ for the proposed WCP constraint. The x and y axes denote the epoch index. The color denotes the value of the components of the matrix. The size of both matrices is $11 \times 11$ where 11 denotes the number of epochs inside the window. In other words, the given satellite is continuously tracked for 11 epochs.

### D. Discussions: Experimental Evaluation in Urban Canyon Using Low-cost Smartphone-level Receiver

The u-blox GNSS receiver and patch antenna belong to the automobile level that is usually used in intelligent transportation (ITS) application. Therefore, it is interesting to see how the proposed method works for the smartphone receiver. The internal antenna of the mobile phone is sensitive to multipath/NLOS and the collected GNSS measurements are noisier. To this end, we collect the other dataset in an urban canyon. The applied smartphone is Huawei P40 Pro and raw GPS/BeiDou measurements are collected at a frequency of 1 Hz. The positioning results of the listed four methods are shown in Table 5. Since the NMEA solution from the smartphone is estimated based on additional information, such as onboard inertial measurement unit, map matching [38], and magnetic sensors, we do not compare the NMEA solution in this experiment. A mean error of 14.45 meters is obtained using the conventional WLS method and the trajectory is shown in Fig. 13 with green curves. After applying the PSR-DOP method, the mean error decreases to 8.74 meters with a decreased standard deviation of 4.21 meters. Interestingly, the mean error increases slightly to 9.19 meters after applying the TDCP constraint. With the help of the proposed method, the mean error decreases to 7.47 meters which shows the effectiveness of the proposed method even in the low-cost smartphone level GNSS receiver. However, we can see that the standard deviation even increases slightly after applying the WCP constraints compared with the PSR-DOP method. One of the major reasons is the unexpected cycle slip in carrier-phase measurements. Due to the limited cost of the GNSS receiver, the cases of the cycle slip are significantly increased compared with the ITS-level u-blox GNSS receiver. Although the Cauchy-based M-estimator can help to mitigate the impacts of the cycle slip, its impacts are still involved in the position estimation problem. It is important to detect the potential cycle slip with the help of other additional onboard sensors, such as IMU to further improve the proposed WCP constraint, which will be one of our future works.

Table 5. Performance of the listed four methods in an urban canyon using smartphone level GNSS receiver

| Item (m) | WLS | Sol 1 | Sol 2 | Sol 3 |
|---|---|---|---|---|
| **MEAN** | 14.45 | 8.74 | 9.19 | 7.47 |
| **STD** | 11.79 | 4.21 | 4.82 | 4.60 |
| **Max** | 110.74 | 46.10 | 45.39 | 46.38 |

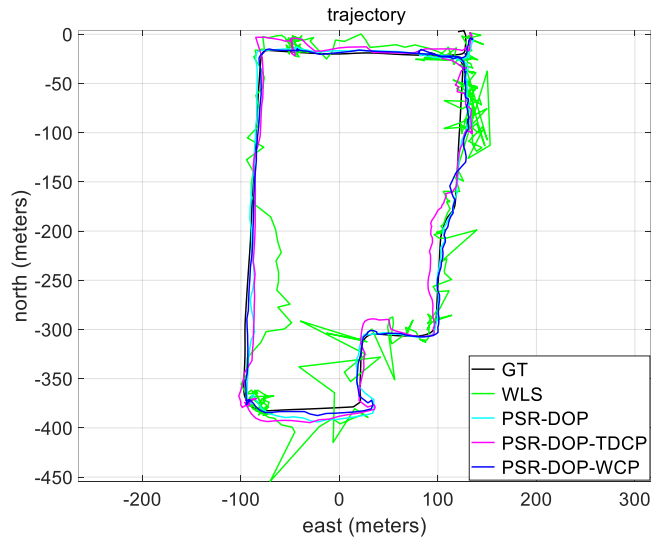

Fig. 13. 2D positioning trajectories of the evaluated four methods in an urban canyon. The black curve denotes the 2D trajectories of the ground truth. The green, cyan, magenta, and blue colors denote the WLS, PSR-DOP (Sol 1), PSR-DOP-TDCP (Sol 2), and PSR-DOP-WCP (Sol 3), respectively.

## VI. CONCLUSIONS

Accurate GNSS positioning in urban canyons is still a challenging problem due to excessive outlier measurements caused by the surrounding high-rising building structures. To improve the urban GNSS positioning, instead of relying on the time difference carrier phase (TDCP) which only considers two neighboring epochs using an extended Kalman filter (EKF) estimator, this paper proposed to employ the carrier-phase measurements inside a window, the so-called window carrier-phase (WCP), to constrain the states inside a factor graph. The effectiveness of the proposed method is verified using both automobile-level and smartphone-level GNSS receivers in typical urban canyons of Hong Kong. One of the major limitations of the proposed method is the reliance on the assumption of constant integer ambiguity for the continuously observed carrier-phase measurements. In the future, we will study the detection of cycle slip by adding the occurrence of the cycle slip as an additional state to be estimated with the help of data redundancy inside the factor graph optimization.

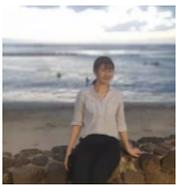

**Xiwei Bai** received an M.S.C. degree in engineering from China Agricultural University in 2018. After that, she worked as a research assistant at Hong Kong Polytechnic University from 2018 to 2019. She is currently studying as a Ph.D. student at the Hong Kong Polytechnic University. Her research interests include visual SLAM and vision-aided GNSS positioning in urban canyons for the intelligent transportation system, autonomous driving.

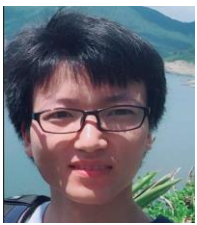

**Weisong Wen** was born in Ganzhou, Jiangxi, China. He is a senior research fellow with the Department of Aeronautical and Aviation Engineering, at Hong Kong Polytechnic University. His research interests include multi-sensor integrated localization for autonomous vehicles, SLAM, and GNSS positioning in urban canyons. He was a visiting student researcher at the University of California, Berkeley (UCB) in 2018.

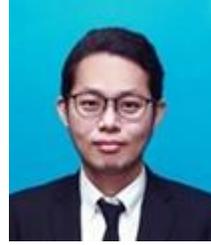

**Li-Ta Hsu** received the B.S. and Ph.D. degrees in aeronautics and astronautics from National Cheng Kung University, Taiwan, in 2007 and 2013, respectively. He is currently an assistant professor with the Interdisciplinary Division of Aeronautical and Aviation Engineering, The Hong Kong Polytechnic University, before he served as a post-doctoral researcher in the Institute of Industrial Science at the University of Tokyo, Japan. In 2012, he was a visiting scholar at University College London, the U.K. His research interests include GNSS positioning in challenging environments and localization for pedestrian, autonomous driving vehicle, and unmanned aerial vehicle.

APPENDIX: CONSTRUCTION OF THE LEFT NULL SPACE MATRIX

### A. Construction of the Left Null Space Matrix

Given a vector $\mathbf{v}_{r,k}^s = \begin{bmatrix} 1 & 1 & ... & 1 \end{bmatrix}^{\mathrm{T}}$, there are several candidate matrices $\mathbf{G}_{r,k}^s$ that satisfying the following equation:

$$\mathbf{G}_{r,k}^s \begin{bmatrix} 1 \\ 1 \\ ... \\ 1 \end{bmatrix} = \begin{bmatrix} 0 \\ 0 \\ ... \\ 0 \end{bmatrix} \tag{A-1}$$

Theoretically, the $\mathbf{G}_{r,k}^s$ encodes the time-correlation of the carrier-phase measurements considered in a WCP by using the equation (15). Different selection of $\mathbf{G}_{r,k}^s$ is corresponding to different confidence levels of time correlation across different epochs of carrier-phase measurements. However, it is still an open question that how to select the optimal $\mathbf{G}_{r,k}^s$ which can optimally capture the time-correlation inside the WCP.

According to our evaluation in Section V, the performance of the WCP is mainly dominated by the size of the window ($N_k^s$) and the quality of the covariance estimation ($\boldsymbol{\Sigma}_{r,k}^{WCP,s}$). Therefore, we adopt the random process [39] to construct the $\mathbf{G}_{r,k}^s$ based on a given size of the window denoted by $N_k^s$. Firstly, the orthonormal basis [39] of the null space of the matrix $\mathbf{v}$ can be calculated which is denoted by matrix $\mathbf{S}_{r,k}^s$. Then a complex matrix $\mathbf{H}_{r,k}^s$ can be constructed by randomly generating two matrices $\mathbf{H}_{r,k,1}^s$ and $\mathbf{H}_{r,k,2}^s$ which making up the real and imaginary part of matrix $\mathbf{H}_{r,k}^s = [\mathbf{H}_{r,k,1}^s \quad \mathbf{H}_{r,k,2}^s]$. Meanwhile, all the components of the two matrices $\mathbf{H}_{r,k,1}^s$ and $\mathbf{H}_{r,k,2}^s$ are ranging between [0,1]. Then the QR factorization [39] is performed as follows:

$$\mathbf{Q}_{r,k}^s \mathbf{R}_{r,k}^s = [\mathbf{H}_{r,k}^s - \mathbf{I}_{r,k}^s] \tag{A-2}$$

where the matrix $\mathbf{Q}_{r,k}^s$ is an orthogonal matrix and the matrix $\mathbf{R}_{r,k}^s$ is an upper triangular matrix. Finally, a unitary matrix can be generated as follows:

$$\mathbf{U}_{r,k}^s = \mathbf{S}_{r,k}^s \mathbf{Q}_{r,k}^s {\mathbf{S}_{r,k}^s}^{\mathrm{T}} + \frac{1}{N_k^s} \mathbf{v}_{r,k}^s {\mathbf{v}_{r,k}^s}^{\mathrm{T}} \tag{A-3}$$

where the $\mathbf{U}_{r,k}^s$ is a complex and unitary matrix. Finally, the $\mathbf{G}_{r,k}^s$ is the imaginary part of the matrix $\mathbf{U}_{r,k}^s$.

### B. *Huber-based Robust Function for WCP Constraint*

Given the error function of the WCP constraint as $\mathbf{e}_{r,k}^{WCP,s}$, the robustified Huber based error function can be formulated as follows:

$$\rho\big(\mathbf{e}_{r,k}^{WCP,s}\big) = \begin{cases} \frac{1}{2}(c)^2, & |\mathbf{e}_{r,k}^{WCP,s}| \le k_\rho \\ k_\rho \left( |\mathbf{e}_{r,k}^{WCP,s}| - \frac{1}{2} k_\rho \right), & |\mathbf{e}_{r,k}^{WCP,s}| > k_\rho \end{cases} \tag{B-1}$$

where the $k_\rho$ denotes the kernel value (parameter).